\documentclass{article}
\usepackage{graphicx} 
\usepackage{graphics}

\usepackage{amsmath}   
\usepackage{amsfonts}
\usepackage{amssymb}

\renewcommand\baselinestretch 2
\begin{document}
%
%
\def\l{\left}
\def\r{\right}
\def\beq{\begin{equation}}
\def\eeq{\end{equation}}
\def\d{\partial}

\newcommand{\W}{{\sf \bfseries {\Large [}Was:~{\Large [}}}
\newcommand{\N}{{\sf \bfseries {\Large ]}Now:~{\Large [}}}
\newcommand{\I}{{\sf \bfseries {\Large [}Insert:~{\Large [}}}
\newcommand{\R}{{\sf \bfseries {\Large [}Remove:~{\Large [}}}
\newcommand{\C}{{\sf \bfseries {\Large [}Comment:~{\Large [}}\sf}
\newcommand{\F}{\normalfont {\sf \bfseries {\Large ]]} }}
\newcommand{\LN}{{\bf (LN:)} }
\newcommand{\mn}{{\bf (MN:)} }

\title{Emergence of complex and spinor wave functions in scale relativity. I. Nature of scale variables}
\author{Laurent Nottale and Marie-No\"elle C\'el\'erier, \footnote{
LUTH, Observatoire de Paris, CNRS, Universit\'e Paris-Diderot,
5 place Jules Janssen, 92195 Meudon Cedex, France;
e-mails: laurent.nottale@obspm.fr, marie-noelle.celerier@obspm.fr}}

\maketitle

\abstract

One of the main results of Scale Relativity as regards the foundation of quantum mechanics is its explanation of the origin of the complex nature of the wave function. The Scale Relativity theory introduces an explicit dependence of physical quantities on scale variables, founding itself on the theorem according to which a continuous and non-differentiable space-time is fractal (i.e., scale-divergent). In the present paper, the nature of the scale variables and their relations to resolutions and differential elements are specified in the non-relativistic case (fractal space). We show that, owing to the scale-dependence which it induces, non-differentiability involves a fundamental two-valuedness of the mean derivatives. Since, in the scale relativity framework, the wave function is a manifestation of the velocity field of  fractal space-time geodesics, the two-valuedness of velocities leads to write them in terms of complex numbers, and yields therefore the complex nature of the wave function, from which the usual expression of the Schr\"odinger equation can be derived.




\section{Introduction}
\label{int}


From a physical point of view, the Scale Relativity theory is the generalization to scales of the relativity principle of motion which underlies the foundation of a large part of classical physics. From a mathematical point of view, it is the giving up of the hypothesis of space-(time) differentiability. Both generalizations result in the fractal nature of space-(time) \cite{GO83,NS84}.

One of this theory main achievements is the foundation of quantum mechanics on first principles \cite{NC07}. In its framework, the quantum mechanical postulates have been derived and the complex, then spinorial, then bi-spinorial nature of the wave function has been naturally recovered \cite{NC07,LN93,CN06,CN03,CN04}, while the corresponding quantum mechanical motion equations, the Schr\"odinger \cite{LN93,JC03}, Pauli \cite{CN06}, Klein-Gordon \cite{LN94,LN96} and Dirac \cite{CN03,CN04} equations have been demonstrated.

The theory also allows one to generalize the quantum mechanical motion equations to the macroscopic realm. This is obtained when the constant ${\cal D}$ which characterizes the amplitude of the fractal fluctuations appearing in the theory, and which corresponds to ${\cal D}=\hbar/2m$ in the standard quantum theory, is given a more general interpretation in terms of a macroscopic constant whose value is linked to the physical system under study. It is therefore possible to derive a macroscopic Schr\"odinger-like equation with numerous applications in physics and other fields (see \cite{LN11} and references therein) and a macroscopic Dirac-like equation whose non-relativistic limit might tentatively reproduce some turbulent fluid behavior \cite{MNC09}.

The emergence of the complex numbers and their generalizations, the quaternions and the bi-quaternions, is issued from the successive doublings of the velocity fields due to the non-differentiability of the fractal functions representing the space-(time) coordinates. These doublings have been extensively studied in previous works (see, e.g., \cite{LN11} for a recent review).

We wish to give here however a new and more detailed derivation of the way the two-valuedness of the mean velocity field naturally emerges in terms of a (+) and (-) velocity, yielding $V$ and $U$ velocity fields, which are subsequently combined in terms of the complex velocity field ${\cal V} = V- i U$ used in the geodesic equation $\widehat{{\rm d}}{\cal V}/{\rm d} t=0$ from which the quantum mechanical motion equations are derived. Then, we are led to generalize these results to three dimensions, being therefore allowed to recover the complex 3-velocity field needed to obtain the Schr\"odinger equation \cite{LN93}.

In a subsequent paper \cite{CN13}, we will be led to extend these results to the 4-velocity field needed to obtain the Klein-Gordon equation \cite{LN94,LN96}. Then we will apply this two-valuedness to successive other doublings, obtaining thus naturally a new expression for the bi-quaternionic velocity field yielding the bi-spinor and the Dirac equation \cite{CN03,CN04}, from which the spinor and the Pauli equation are derived \cite{CN06}.

The structure of the present paper is as follows. In Sec.~\ref{complex}, we give an improved physically complete picture of how the complex numbers emerge in the quantum mechanical domain from the doubling of the velocity fields in a non-differentiable space. In Sec.~\ref{3D}, we generalize to three dimensions the one-dimensional results previously obtained and in Sec.~\ref{schro}, we recall how the Schr\"odinger equation eventually emerges and then complete the proof of the two-valuedness of all components of the velocity field of the geodesics fluid. In Sec.~\ref{concl}, we present our conclusions.


\section{Emergence of complex numbers in one dimension}
\label{complex}


\subsection{General argument}
\label{genarg}

The first step is the theorem according to which a continuous and non-differentiable curve is fractal, in the general meaning that its length is explicitly scale dependent and tends to infinity when the resolution interval tends to 0 \cite{LN93,BAC00,JC02}. In terms of a parameter $t$ along this curve (which may, in particular, be a time coordinate), and identifying 
a small interval ${\rm d} t$ of the $t$ parameter to the scale variable (i.e., to a continuous resolution interval), it reads $(L=L(t,{\rm d} t) \to \infty)_{{\rm d} t \to 0}$.

\begin{figure}[!ht]
\begin{center}
\includegraphics[width=8cm]{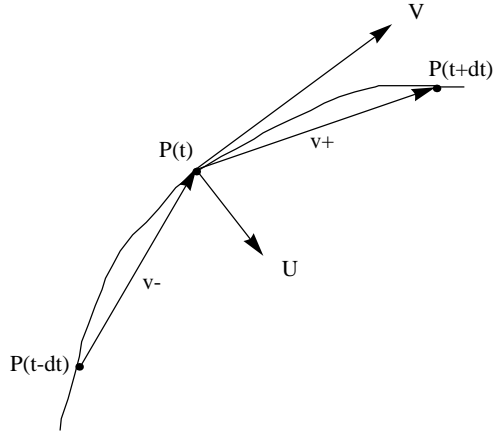}
\caption{\small Rough illustration of the definition of the two velocities at point $P(t)$: $V_+$, defined using a point $P(t+dt)$ which follows $P$, and $V_-$, defined using a point $P(t-dt)$ which precedes (we take here $dt>0$). In the differentiable case, $V_+$ and $V_-$ are defined and $V_+=V_-$ at the limit $dt \to 0$. On the contrary, in the nondifferentiable case,  $V_+$ and $V_-$ remain different for all values of $dt$ and are undefined at the limit $dt \to 0$. From these two velocities, one builds another doublet representation as $V=(v_+ + v_-)/2$ and $U=(v_+ - v_-)/2$, then a complex representation as ${\cal V}=V- i U$.}
\label{fig1b}
\end{center}
\end{figure}

A continuous fractal coordinate therefore reads $X=X(t,{\rm d} t)$ and it is non-differentiable, under the meaning that, though differential elements ${\rm d} X$ and ${\rm d} t$ can be defined as a consequence of continuity, their ratio ${\rm d} X/{\rm d} t \to \infty$ when ${\rm d} t \to 0$. The derivative is therefore redefined as a fractal function \cite{LN93}, i.e., as a function which is explicitly dependent on the scale interval ${\rm d} t$. But this new definition is now two-valued:

\beq
V_+(t,{\rm d} t)=\frac{{\rm d}_+X }{{\rm d} t}= \l(\frac{X(t+{\rm d} t,{\rm d} t)-X(t,{\rm d} t)}{{\rm d} t}\r)_{{\rm d} t > 0}, 
\eeq
\beq
V_-(t,{\rm d} t)=\frac{{\rm d}_-X }{{\rm d} t}= \l(\frac{X(t+{\rm d} t,{\rm d} t)-X(t,{\rm d} t)}{{\rm d} t}\r)_{{\rm d} t < 0},
\eeq
which can be written equivalently as
\beq
V_-(t,{\rm d} t)= \l(\frac{X(t,{\rm d} t)-X(t -{\rm d} t,{\rm d} t)}{{\rm d} t}\r)_{{\rm d} t > 0},
\eeq
\beq
V_+ (t,{\rm d} t) = \l(\frac{X(t+{\rm d} t,{\rm d} t)-X(t,{\rm d} t)}{{\rm d} t}\r)_{{\rm d} t > 0},
\eeq
which are before and after derivatives defined respectively between the points $t - {\rm d} t$ and $t$ and the points $t$ and $t + {\rm d} t$, (see Fig.~\ref{fig1b}).

\begin{figure}[!ht]
\begin{center}
\includegraphics[width=8cm]{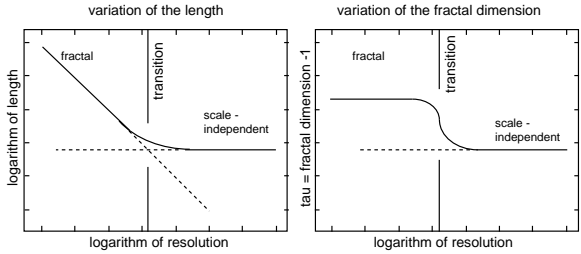}
\caption{\small Example of an explicitly scale-dependent curve: the variation of the length of the curve $L(t,{\rm d} t)$ is shown in terms of the resolution interval ${\rm d} t$ (log-log plot) for a given value of the time $t$. The length tends to infinity when ${\rm d} t$ tends to 0.}
\label{fig1}
\end{center}
\end{figure}

Let us be more specific about the meaning of this new definition. The usual definition of the derivative of a differentiable function $X(t)$ consists in {\em taking the limit}, when ${\rm d}t \to 0$, of $(X(t+{\rm d}t)-X(t))/{\rm d}t$. In the nondifferentiable case considered here, this limit no longer exists, so that this usual definition fails. The new definition is based on the remark that nondifferentiability does not prevent one to defining intervals ${\rm d}X$ and ${\rm d}t$, whose existence relies on the continuity of space-time ($X,\, t$). Nor does it prevent to define the ratio of these intervals, ${\rm d}X/{\rm d}t$, as long as they remain finite, i.e. ${\rm d}t \neq 0$. It is only the limit ${\rm d}t \to 0$ which is no longer defined. So the above definition means that one considers all possible values of $[X(t+{\rm d} t,{\rm d} t)-X(t,{\rm d} t)]/{{\rm d} t}$ and of ${\rm d}t$ when ${\rm d}t \to 0$, and keeps the information about these values in terms of a function which is explicitly dependent on ${\rm d}t$, considered as a full variable as can be seen in Fig.~\ref{fig1}. Therefore the nondifferentiable case just corresponds to the case when this function is not defined for the value ${\rm d}t=0$ (note that, in terms of the more relevant scale variable $\ln ({\rm d}t/ \tau)$, this limit becomes  $\ln ({\rm d}t/ \tau) \to -\infty$). 

Here, $\tau$ is a reference resolution needed since only resolution ratios, not resolutions, have a physical meaning. It fixes arbitrarily the units in which the resolutions are computed and is needed to ensure that the quantity in the logarithm is dimensionless. 

This generalized definition of the derivative includes the usual differential one, in which one takes the limit ${\rm d} t \to 0$, as a particular case. Indeed, if the limit exists, as in the differentiable case, then it is included in this definition as $V(t)=V_+(t,0)=V_-(t,0)$. If it does not exist, the usual definition fails, since the derivative is in this case undefined, while the new definition, which includes all the history of the way the function behaves when ${\rm d} t \to 0$ (including its possible divergence) is effective as a tool with which one can effectively work.

Let us give a simple example of the possible consequences and usefulness of this enlarged definition in physics. Consider a physical function $A$ which could be shown to be vanishing, e.g. $A(t,{\rm d}t)=t {\rm d}t$. Under the usual definition, where one takes the limit ${\rm d}t \to 0$, one would have $A=0$. Assume now that another physical quantity is divergent, i.e., $B(t,{\rm d}t)= t/{\rm d}t$ (this is for example the case for the mean square value of velocity in quantum mechanics \cite{FH65}). In terms of the usual calculus, one would claim that this function is undefined when ${\rm d}t \rightarrow 0$. But if a new physical quantity writes $C=A B$, the usual approach would conclude that it is undefined, $C=0 \times \infty$, while the new definition where the differential elements are kept as explicit variables yields a finite function of time, $C(t)= t^2 $.

In terms of the above definition, in which ${\rm d} t$ is allowed to be positive or negative, there is, strictly, only one function, $V(t,{\rm d} t)$. The two-valuedness is a manifestation of the fact that, in general, there is no reason why it should be symmetric with respect to the variable ${\rm d} t$. However, from the view point of scale laws, i.e., when looking to a specific transformation ${\rm d} t \to {\rm d} t'$, the natural scale variable is not ${\rm d} t$ but its logarithm, which acts as a kind of theoretical magnifying glass to view the behavior of the function around ${\rm d} t=0$. 

\begin{figure}[!ht]
\begin{center}
\includegraphics[width=8cm]{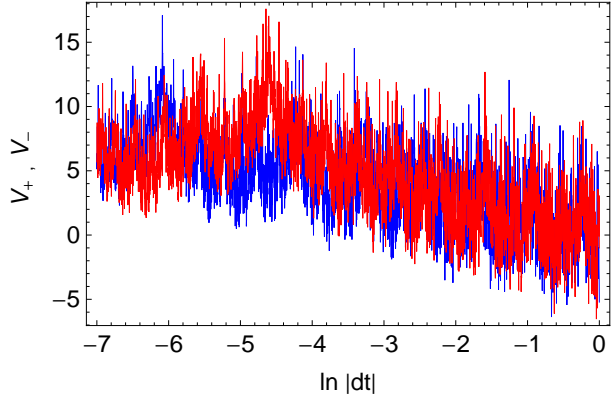}
\caption{\small Example of the two-valuedness of the derivative of a non-differentiable and fractal function. The function in this example is a Weierstrass function, given here by $X(t)=\sum_{k=0}^\infty \sin(2^k t)/2^k$. Thererefore its time derivative is $V(t)=dX/dt=\sum_{k=0}^\infty \cos(2^k t)$. The figure gives the derivative at the point $t_0= 0.01$ in terms of the logarithm of a small interval $dt=t-t_0$. Since one should take the absolute value of $dt$ in the logarithm, two derivatives must be defined, before the point $t_0$ ($t<t_0$, blue curve) and after this point ($t>t_0$, red curve). It is clearly apparent that the two derivatives, which are continuous functions of $\ln|dt|$, are not given by the same function.}
\label{fig2}
\end{center}
\end{figure}

This implies to jump to its absolute value $|{\rm d} t|$ and to choose a reference scale $\tau$ in order to write it as $\rho_t=\ln(|{\rm d} t|/\tau)$. This writing is just a manifestation of the principle of scale relativity, according to which scales do not exist as such, but only through their ratios. If we want now to plot the function $V(t,\rho_t)$ , it is  clear that we are obliged to use two functions, $V_+=[V(t,\rho_t)]_{{\rm d} t>0}$ and $V_-=[V(t,\rho_t)]_{{\rm d} t<0}$ (see Fig.~\ref{fig2}).

Another result of the scale relativity approach is that such a scale-dependent fractal function can be generally written as the sum of a differentiable ``classical" part and of a non-differentiable, divergent fractal part. In the case of a fractal velocity field, this reads
\beq
V_+[x(t,{\rm d} t),t,{\rm d} t]=v_+[x(t),t]+w_+[x(t,{\rm d} t),t,{\rm d} t] ,
\eeq
\beq
V_-[x(t,{\rm d} t),t,{\rm d} t]=v_-[x(t),t]+w_-[x(t,{\rm d} t),t,{\rm d} t].
\eeq
where  $w_\pm = {\rm d} \xi_\pm/{\rm d} t$ and $|{\rm d} \xi_{\pm}|^{D_f} \propto  |{\rm d} t|$, $D_f$ being the fractal dimension of the geodesics.

In previous publications, we have considered that, in general, there is no a priori reason for the ``classical" velocity fields $v_+$ and $v_-$ to be the same. It is this specific point that we want to address and to elaborate here, in order to understand better the reason why this two-valuedness affects not only the fractal (scale-dependent) part of the velocity, but also its ``classical", scale-independent, differentiable part.

\subsection{Two-valuedness of the mean velocity field}
\label{twoval}

Let us consider a fractal coordinate $X(t, \epsilon)$. By ``fractal'', we mean that it is explicitly dependent on a scale variable (or ``resolution'')  $\varepsilon>0$ and divergent when $\varepsilon \to 0$. Being defined as a resolution, $\varepsilon$ is fundamentally positive, i.e., $\varepsilon \in {\mathbb{R}}+$.
This resolution may be of two types, time-like ($\varepsilon_t$) or space-like ($\varepsilon_X$). The relation between the time-resolution and space-resolution on a fractal curve of fractal dimension $D_f$ is:
\beq
\varepsilon_X^{D_f} \propto \varepsilon_t.
\eeq

Consider the case when the curve described by the curvilinear coordinate $X$ is traveled during time $t$. In terms of the time-resolution, the coordinate then reads $X= X(t,\varepsilon_t)$. For any given value of $\varepsilon_t$, the curve is smooth and differentiable.

\begin{figure}[!ht]
\begin{center}
\includegraphics[width=8cm]{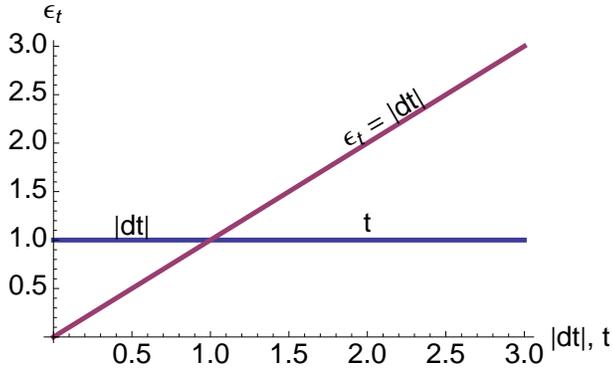}
\caption{\small Time measurements on a  physical "object" and their theoretical description involve the measurement resolution $\varepsilon_t$, the time variable $t$ and the time differential element ${\rm d} t$. The same is true for space $x$ and other measurements. Values of measurement results are by construction $>\varepsilon_t$ and are represented by a time variable $t$. Differential elements used in a theoretical description are therefore such that ${\rm d} t \leq \varepsilon_t$. When the object is explicitly resolution-dependent (``fractal'' in a general meaning), it changes with the value of $\varepsilon_t$, so that the differential element ${\rm d} t$ is no longer defined on the same object. A solution to this problem consists in placing oneself on the resolution "interface", e.g., to work with differential elements equal to the resolution, i.e., such that $|{\rm d} t|=\varepsilon_t$.}
\label{figa}
\end{center}
\end{figure}

It is also quite possible to define a time differential element ${\rm d} t \to 0$ on this curve.  The time measurement resolution $\varepsilon_t$ defines the transition between what is considered as the time variable $t>\varepsilon_t$, which is measured in units of this resolution and the differential element ${\rm d} t \leq \varepsilon_t$ tending to 0 (see Fig.~\ref{figa}).

For this curve, observed, measured or considered at a fixed resolution $\varepsilon_t$, we may then define a mathematical velocity in the usual differentiable way:
\beq
\tilde{V}_{\varepsilon_t}(t)= \frac{{\rm d} X_{\varepsilon_t}}{{\rm d} t}={\rm lim}_{{\rm d}t\to 0}\l( \frac{X_{\varepsilon_t}(t+{\rm d} t)-X_{\varepsilon_t}(t)}{{\rm d} t}\r).
\eeq
We have written here ${\varepsilon_t}$ as an index, since it is fixed at a given value, instead of being a variable. It must be understood that this derivative is a purely mathematical tool, valid only on the curve of resolution $\varepsilon_t$ and allowing to make differential calculus on this curve, but that it does not represent the true velocity on the fractal curve. Indeed, when the time resolution interval is decreased ($\varepsilon'_t < \varepsilon_t$), completely new information appears on the fractal curve, which was unpredictable from the sole knowledge of $ X_{\varepsilon_t}$. A new velocity $\tilde{V}_{\varepsilon'_t}(t)$ can now be defined, which may be fundamentally different from $\tilde{V}_{\varepsilon_t}(t)$ . This derivative has only mathematical meaning at scales smaller than ${\varepsilon'_t}$, and physical meaning at scale ${\varepsilon'_t}$.

Considering again ${\varepsilon_t}$ as an explicit variable, this means that one can define a ``fractal velocity" as
\beq
V_f(t,{\rm d} t,\varepsilon_t)=\frac{{\rm d}X }{{\rm d} t}=\frac{X(t+{\rm d} t,\varepsilon_t)-X(t,\varepsilon_t)}{{\rm d} t},
\eeq 
where $\varepsilon_t >0$, while ${\rm d} t$ is algebraic and can be positive or negative, i.e., $ {\rm d} t \in {\mathbb{R}}$. 

It is now clear from the above discussion that there is a particularly meaningful choice for the scale variable, which is $\varepsilon_t = |{\rm d} t|$. This special choice  amounts to place oneself just on the interface where the derivative takes its physical value whatever the scale. This choice corresponds to the natural definition of a resolution from the theoretical viewpoint, in accordance with the Riemann-Lebesgue method (while from the experimental viewpoint, resolutions can be intervals like in a Charpak multi-wire detector, but may also be defined as pixels, or covering balls, standard errors in statistical measurements, etc., see \cite[Chap. 3]{LN11}). The interval ${\rm d} t$ has therefore two different and complementary roles: (1) as a differential element, it appears in sums like $t + {\rm d} t$ and is algebraic; (2) as a resolution interval, its sign looses its meaning and it appears in scale transformations in a logarithmic form, such as $\ln(|{\rm d} t|/\tau)$.

More generally, in a fractal space, we are led to consider a fractal velocity field $V(x(t),t,{\rm d} t,\varepsilon_t) ={\rm d}X /{\rm d} t$, so that the differential element ${\rm d} X(x(t),t; {\rm d} t,\varepsilon_t)$ can be decomposed in terms of various contributions, a differentiable linear one and a non-linear fractal fluctuation:
\beq
{\rm d}X = V[x(t),t] \,{\rm d} t +  U[x(t),t] \, \varepsilon_t + {\rm d} \xi
\label{8}
\eeq
The fractal fluctuation can be described by a stochastic variable which has the nature of a space resolution  $\varepsilon_x>0$, i.e.,
\beq
{\rm d} \xi= \eta \: \varepsilon_x,
\eeq
where $\eta$ is a purely mathematical dimensionless variable which is normalized according to $<\eta>=0$ and $<\eta^2>=1$ and may have any probability distribution.
When the fractal dimension is $D_f=2$, the space and time resolutions are related by
\beq
\varepsilon_x = \sqrt{2 {\cal D} \varepsilon_t}.
\eeq
We obtain therefore
\beq
{\rm d}X = V[x(t),t] \, {\rm d} t +  U[x(t),t] \, \varepsilon_t +  \eta \,  \sqrt{2 {\cal D} \varepsilon_t}.
\eeq
If we place ourselves now on the interface $\varepsilon_t = |{\rm d} t|$ which defines the physical fractal derivative, we have
\beq
{\rm d}X = V {\rm d} t +  U |{\rm d} t| +  \eta \sqrt{2 {\cal D} |{\rm d} t|}.
\eeq
Therefore two possibilities occur for the elementary displacements:
\beq
{\rm d} t>0: \;\;\; |{\rm d} t|={\rm d} t, \;\;\; {\rm d}_+X = (V +  U) \, {\rm d} t +  \eta \,  \sqrt{2 {\cal D} {\rm d} t}.
\eeq
\beq
{\rm d} t<0: \;\;\; |{\rm d} t|=-{\rm d} t, \;\;\; {\rm d}_-X = (V - U) \, {\rm d} t +  \eta  \, \sqrt{-2 {\cal D} {\rm d} t}.
\eeq
By setting $v_+=V+U$ and $v_-=V-U$, we recover the two-valuedness of the mean velocity field in terms of a (+) and (-) velocity. 

In this new and more detailed derivation, it is noticeable that the V and U velocity fields appear first, while they were derived from the v(+) and v(-) velocities in the previous ones.

The next step of the construction of the scale relativity theory consists of combining the V and U velocity fields into a complex velocity field:
\begin{equation}
{\cal V} = V - i U.
\end{equation}
The question of the choice of complex numbers to represent this two-valuedness of velocity has been already fully addressed in \cite{LN08} and \cite{LN11} Sec. 5.4.1. Let us summarize the argument. 

Each real component of the velocity is now replaced by a doublet of real numbers. In such a case, the sum operation is easy to generalize, but one needs to define a new product. From the mathematical point of view, this is the well-known problem of the doubling of algebra (see, e.g.,  \cite{MP82}). The mathematical solution to this problem \cite{EC53} is precisely that the doubling $\mathbb{R}^2$ of real numbers $\mathbb{R}$ is the algebra $\mathbb{C}$ of complex numbers, the doubling $\mathbb{C}^2$ of $\mathbb{C}$ is the algebra $\mathbb{H}$ of quaternions and the doubling $\mathbb{H}^2$ of quaternions is the algebra of Graves-Cayley octonions. Recall that these successive algebra doublings lead to successive deterioration of the algebraic properties (namely, loss of order relation in complex plane, then of commutativity for quaternions, then of associativity for octonions), and that the Cartan solution is the optimal one in this respect. 

One reaches the same conclusion with a physical argument which is specific of the present situation. One may write the velocity doublet $(V,U)$ under the form $ {\cal V} = V + \alpha U$ by identifying the doublet $(V,0)$ with the real velocity $V$. Then the knowledge of the new product that we want to define can be reduced to the mere knowledge of the square $\alpha^2$. The Lagrange function of a free particle, when it is calculated using the full velocity field, i.e. including the fractal diverging part ${\cal W} = \left(\frac{w_{+}+w_{-}}{2}-\alpha\, \frac{w_{+}-w_{-}}{2}\right)$, reads ${\cal L} = \frac{1}{2} m ({\cal V}^2 + <\!{\cal W}^2\!>$). But $< \!{\cal W}^{2}\! >  =\frac{1}{4}< (w_{+}^2+w_{-}^2)(1+\alpha^2)>$, and then the choice $1+\alpha^2=0$, i.e. the choice of the complex numbers $\alpha= \pm i$, suppresses an infinite term in the theory and reduces the full Lagrange function to its previous simple expression in terms of the only mean velocity field, ${\cal L} = \frac{1}{2} m {\cal V}^2$. Therefore complex numbers achieve a physically highly significant choice of representation of the two-valuedness of velocities, that can be called ``covariant" in a general meaning of this word, i.e., a representation in which the equations keep their previous (simplest) form. 

The fields $V$ and $U$ acquire here a new status, $V$ being the component linked to the algebraic differential element ${\rm d} t$ while $U$ is the component linked to $|{\rm d} t|$, now understood as a scale variable having the properties of a resolution interval submitted to scale transformations $\ln(|{\rm d} t|/\tau) \to \ln(|{\rm d} t'|/\tau)$. This result reinforces the choice of identifying $V= (v_++v_-)/2$ with the real part of the complex velocity. Indeed, in the new writing, ${\rm d} X=V {\rm d} t$ is the term obtained in usual differentiable calculus, while the additional term $U \varepsilon_t$ is new and linked to the explicit scale dependence. Moreover, $V[x(t),t]$ is the velocity field that naturally appears in the fluid mechanics form of the Schr\"odinger equation (continuity equation + Euler equation including a quantum potential \cite{LN09}) (see Sec.~\ref{schro}).


\section{Generalization to three dimensions of the one-dimensional results}
\label{3D}


To derive the Schr\"odinger equation, a generalization of the one-dimensional results obtained above to the three dimensional space is needed. For each of the three coordinates defining this space, Eq.~(\ref{8}) can be written as
\begin{equation}
{\rm d} X_i = V_i[x_i(t),t] {\rm d} t + U_i[x_i(t),t] \varepsilon_t + {\rm d} \xi_i,
\label{dXi}
\end{equation}
with $i = 1,2,3$, ${\rm d} \xi_i = \eta_i \varepsilon_{x_i}$, $\varepsilon_{x_i} = \sqrt{2 {\cal D} \varepsilon_t}$,  and the dimensionless variables $\eta_i$ being normalized according to $\langle \eta_i \rangle = 0$, $\langle \eta_i \eta_j \rangle = \delta_{ij}$, where $\delta_{ij}$ is the Kronecker symbol. This implies
\begin{equation}
{\rm d} \xi_i = \eta_i \sqrt{2 {\cal D} \varepsilon_t},
\label{dxi1}
\end{equation}
which, with the particularly meaningful choice $\varepsilon_t = |{\rm d} t|$, gives
\begin{equation}
{\rm d} \xi_i = \eta_i \sqrt{2 {\cal D} |{\rm d} t|}.
\label{dxi2}
\end{equation}
The two possibilities for the elementary displacements are therefore
\begin{equation}
{\rm d} t > 0 \qquad |{\rm d} t| = {\rm d} t, \qquad {\rm d} \xi_{i+} = \eta_i \sqrt{2 {\cal D} {\rm d} t},
\label{dxiplus}
\end{equation}
and
\begin{equation}
{\rm d} t < 0 \qquad |{\rm d} t| = - {\rm d} t, \qquad {\rm d} \xi_{i-} = \eta_i \sqrt{- 2 {\cal D} {\rm d} t},
\label{dximoins}
\end{equation}
which gives, for the expectation value of the product of two independent fractal fluctuation fields,
\begin{equation}
\langle {\rm d} \xi_{i \pm} {\rm d} \xi_{j \pm} \rangle = \pm 2 \delta_{ij} {\cal D} {\rm d} t.
\label{dxipm1}
\end{equation}
This shows how Eq.~(\ref{dxipm1}), from which the derivation of the Schr\"odinger equation proceeds \cite{LN93,LN11}, can be obtained naturally with the above particularly meaningful choice for the scale variable.


\section{Derivation of the Schr\"odinger equation in Scale Relativity: a reminder}
\label{schro}


The theory of scale relativity is the extension of the relativity principle of motion to scale transformations. In this framework, the quantum properties of a particle are the manifestations of the fractal structure of space(-time). In particular, the complex nature of the wave function and the Schr\"odinger equation as a geodesic equation in a fractal space emerge naturally. We have discussed at length the complex nature of the wave function in Sec.~\ref{complex}. Now we give below a reminder of the way the Schr\"odinger equation is obtained.

The basic principle is a relativity principle. The laws of physics must be such that they apply whatever the reference system state, which means that physical quantities are not defined in an absolute way but are relative to the state of the reference system. This principle is implemented by three related principles provided with the corresponding mathematical tools.

The covariance principle states that the equations of physics keep the same simplest form under changes of the reference system state. It can be strong, such as the motion equations in General Relativity or weak, such as the Einstein field equations with source terms.

The equivalence principle is a more specific statement of the relativity principle applied to a given physical domain. In General Relativity, a gravitational field is locally equivalent to an acceleration field. Hence, it exists coordinate systems in which gravitation locally disappears. Similar proposals apply in Scale Relativity: a quantum behavior is locally equivalent to fractal (i.e., non-differentiable) motion, gauge fields are locally equivalent to internal resolution transformations.

The geodesic principle states that free trajectories are space-time geodesics. Therefore the dynamics equations are determined by the space-time geometry. The action element identifies with the metric invariant, i. e., the proper-time, $dS = -mc \; ds$ which implies that the action principle is equivalent to a geodesic principle.

A covariant derivative is the main mathematical tool by which the above principles are implemented.
This new derivative includes all the effects of the geometry, which is at variance with usual field theories where they are externally applied to the system.

In General Relativity, the geometric effects are subtracted from the total increase of, say, a vector, leaving only the inertial part which defines the covariant derivative as
\beq
D_{\mu}V_{\nu} = \partial_{\mu}V_{\nu} - \Gamma^{\rho}_{\mu \nu} V_{\rho}.
\label{codegr}
\eeq

The most remarkable result is that the three principles (strong covariance, equivalence and geodesic principles) lead to the same form of the motion equations, i. e., the Galileo form for the inertial motion of a free body, but written with the covariant derivative as
\beq
Du_{\mu}/ds = 0.
\label{galileomot}
\eeq

In Scale Relativity, a similar construction is used and a covariant derivative is constructed in order to implement similar principles, here applied to the relativity of scales.

Now, in physics, any measurement is made with a finite resolution $\epsilon$. However, the differential equations describing the laws are written and solved as if the differential elements ${\rm d}x$, ${\rm d}t$, ${\rm d}s$ etc. tend to zero, eventually reaching the null value.

As seen in Sec.~\ref{complex}, in Scale Relativity, the mathematical tools ${\rm d}x$, ${\rm d}t$, ${\rm d}s$ are assimilated to physical resolutions always non vanishing, which implies that null differential elements are actually physically impossible (since, from quantum mechanics, an infinite energy-momentum would be needed to effectively realize them). Moreover, a physical quantity $f=f(x)$ where $x$ is a space-time variables is also more generally depending also on resolutions, i. e., $f=f(x,\epsilon)$, more precisely $f=f(x, \ln \epsilon)$ \cite{LN93,LN11}.

Thus, for $\epsilon = 0$, the function $f(x)=f(x,-\infty)$ defined at the limit ${\rm d}x \to 0$ is devoid of physical meaning (in the nondifferentiable case).

To show that $f$ is indeed dependent not on the resolution but on its logarithm, one can use the Gell-Mann-Levy method and apply an infinitesimal dilation, $\epsilon \rightarrow \epsilon' = \epsilon (1 + {\rm d} \rho)$, to the resolution. The length, ${\cal L}$ of a fractal curve is thus, to first order,
\beq
{\cal L}(\epsilon') = {\cal L}(\epsilon) + \epsilon \, {\rm d} \rho \; \partial {\cal L}(\epsilon) / \partial \epsilon,
\label{Gell1}
\eeq
which can be written as
\beq
 {\cal L}(\epsilon') = (1 + {\tilde D} \; {\rm d} \rho) {\cal L}(\epsilon),
\label{Gell2}
\eeq
with
\beq
{\tilde D} = \epsilon \; \frac{\partial}{\partial \epsilon} = \frac{\partial}{\partial  \ln \epsilon}. \label{Gel3}
\eeq

This form of the infinitesimal dilation operator, ${\tilde D}$, shows that the natural variable for the resolutions is $\ln \epsilon$. Then fractal functions are considered, $f(x,\epsilon)$ (here and in the following, we write, for simplicity, the dependence on $\ln \epsilon$ as a dependence on $\epsilon$), differentiable for all $\epsilon \neq 0$, which means that all scales coexist in a `scale space' and are connected together via scale differential equations.

The simplest pure scale law is first order and states that the variation of ${\cal L}$ under an infinitesimal ${\rm d} \ln \epsilon$ depends only on ${\cal L}$. It reads therefore:
\beq
\frac{\partial {\cal L}(s,\epsilon)}{\partial \ln \epsilon} = \beta({\cal L}) = a + b {\cal L}.
\label{simplescalelaw}
\eeq

The solution of this renormalization group-like equation is a fractal length which can be written as
\beq
{\cal L}(s, \epsilon) = {\cal L}_0(s)\left[1 + \zeta(s)\left(\frac{\lambda} {\epsilon}\right)^{\tau}\right],
\label{renormeq}
\eeq
whose projection on the X axis is
\beq
X(s, \epsilon) = x(s)\left[1 + \zeta_x(s)\left(\frac{\lambda}{\epsilon}\right)^{\tau}\right],
\label{projrenormeq}
\eeq
with $\zeta_x(s)$, a highly fluctuating (fractal) function possibly described by a stochastic variable, and with $\tau = -b$.

The simplest case which can now be considered is a constant fractal dimension $D_F=2$. The reasons for making it conspicuous are: (i) Feynman showed that typical quantum mechanical paths, contributing most to the path integral, are non-differentiable and of fractal dimension 2 \cite{FH65}; (ii) a fractal dimension 2 is that of Brownian motion, more generally of Markov processes, typical of uncorrelated motion reproducing the effect of a fractal space(-time) on the elementary displacements ${\rm d} \xi$;
(iii) it is only in the case $D_F=2$ that the explicit scale dependent terms in the generalized Schr\"odinger equation disappear, therefore allowing us to recover the usual form of the Schr\"odinger equation; (iv) it is possible to define a finite internal angular momentum (quantum spin) for a fractal spiral path with fractal dimension 2. For $D_F<2$ it is null, for $D_F>2$ it is infinite \cite{LN11}.

Let us continue the derivation in the simple case accounted for in the present paper: non-differentiability of the three-dimensional space and breaking of the reflection invariance of the time differential element ${\rm d} t \leftrightarrow \, - \, {\rm d} t$ (see \cite{CN13} for a more general case). The derivative with respect to time of a differentiable function $f(t)$ is
\beq
\frac{{\rm d} f}{{\rm d} t} = \lim_{{\rm d}t \to 0, {\rm d}t >0} \frac{f(t+{\rm d}t) - f(t)}{{\rm d}t} =  \lim_{{\rm d}t \to 0, {\rm d}t <0} \frac{f(t) - f(t-{\rm d}t)}{{\rm d}t}
\label{derivdifffunc}
\eeq
In the non-differentiable case the limits are no longer defined, hence both definitions fail.

As explained in detail in Sec.~\ref{complex}, a solution is to replace $f(t)$ by an explicitly scale-dependent fractal function $f(t,{\rm d} t )$ of two variables: $t$ in space(-time) and ${\rm d} t$ in scale space. Then two functions can be defined as
\beq
f'_+(t, {\rm d}t) = \frac{f(t + {\rm d}t,{\rm d}t)-f(t,{\rm d}t)}{{\rm d}t},
\label{fplus}
\eeq
\beq
f'_-(t,{\rm d}t) = \frac{f(t,{\rm d}t)-f(t-{\rm d}t,{\rm d}t)}{{\rm d}t}.
\label{fmoins}
\eeq
It is easy to see from Eqs.~(\ref{fplus}) and (\ref{fmoins}), that $f'_+ \leftrightarrow  f'_-$ is equivalent to ${\rm d}t \leftrightarrow - \, {\rm d}t$. Thus the differential time reflection, which is a discrete symmetry of standard physics, is broken.

Let us give an additional argument illustrating the two-valuedness of the derivative in a scale-dependant situation. Consider a function $X(t)$ and its increment ${\rm d}X$. By definition,
\beq
\frac{{\rm d}X}{{\rm d}t}= \frac{X(t+{\rm d}t)-X(t)}{{\rm d}t},
\eeq
For ${\rm d}t \ll t$ the Taylor expansion of $X(t+{\rm d}t)$ is
\beq
X(t+{\rm d}t)=X(t)+ X'(t) \,{\rm d}t + \frac{1}{2} X''(t) \,{\rm d}t^2+...,
\eeq
which gives, for the ratio ${\rm d}X/{\rm d}t$,
\beq
\frac{{\rm d}X}{{\rm d}t}=X'(t) + \frac{1}{2} X''(t) {\rm d}t+...
\eeq
which can be written as ${\rm d}X/{\rm d}t = V+ {\rm d}V/2$. In the usual differential case where $X''(t)$ remains finite, i.e. ${\rm d}V \ll V$, the last term vanishes when ${\rm d}t \to 0$, and one recovers the usual identification between $X'(t)$ and ${\rm lim}_{{\rm d}t \to 0} ({\rm d}X/{\rm d}t)$. But in fractal conditions one can have ${\rm d}V \sim V$, and therefore an additional term is lacking in the usual differential calculus. For example, if $X''(t)=X''(t,{\rm d}t) \propto 1/{\rm d}t$,  $\frac{1}{2} X''(t) {\rm d}t$ becomes finite and must be taken into account in the derivative. 

\begin{figure}[!ht]
\begin{center}
\includegraphics[width=12cm]{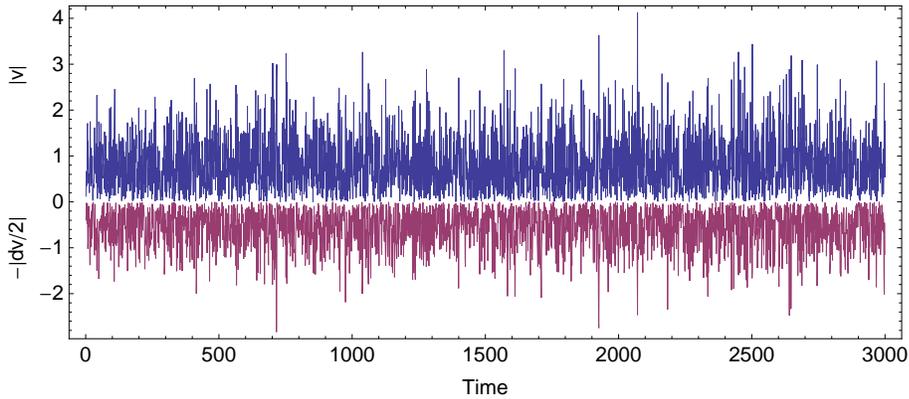}
\caption{{\small Comparison between the absolute values $V_i$ of a Gaussian stochastic variable $V$ with zero mean and dispersion 1 (top curve) and its increments ${\rm d}V = V_{i+1} - V_i$ (bottom).}}
\label{fig4}
\end{center}
\end{figure}

A simple example of such a case is given by a Gaussian stochastic variable $V$ for which we have also computed the increment ${\rm d}V$. The result is given in Fig.~\ref{fig4}, where $V$ and ${\rm d}V$ are indeed found to be of the same order of magnitude. 

A two valuedness of the derivative under the reflexion ${\rm d}t \to -{\rm d}t$ emerges as
\beq
V_+= V+\frac{1}{2} {\rm d}V, \;\;\;   V_-=V-\frac{1}{2} {\rm d}V
\eeq
and therefore 
\beq
V= \frac{V_+ + V_-}{2}, \; \; \; U= \frac{V_+ - V_-}{2}=\frac{1}{2} \, {\rm d}V.
\eeq
The correspondance of the drift velocity $U$ with an acceleration, $\Gamma={\rm d}V/{\rm d}t=2U/{\rm d}t$ was already pointed out in \cite[Chap. 5, p. 142]{LN93}.

Applying these definitions to fractal space coordinates $x(t,{\rm d} t )$, one obtains two velocity fields which can be decomposed into their classical and fractal parts as 
\beq
V _+ [x(t,{\rm d}t), t, {\rm d}t ] = v_+[x(t), t] + w_+[x(t,{\rm d}t), t, {\rm d}t ],
\label{plusvel}
\eeq
\beq
V _- [x(t,{\rm d}t), t, {\rm d}t ] = v_-[x(t), t] + w_-[x(t,{\rm d}t), t, {\rm d}t ].
\label{minusvel}
\eeq
Now, we have seen in Sec.~\ref{complex}, that there is no a priori reason for  $V_+$ and $V_-$ to be the same functions, so that there is a general two-valuedness when accounting for the differentiable part. 

However, at this stage of the derivation, there is no proof that the same is true concerning the  differentiable parts $v_+$ and $v_-$ alone. Clearly, if one considers only, not a fractal velocity field but a given particular fractal curve, there is no doubling for such a single curve, since it is smoothed out at scales larger than the fractal to non-fractal transition, beyond which $w \ll v$. But for a velocity field, the question remains open at this stage. Therefore, we just introduce the doubling for generality and work with the two velocities $v_+$ and $v_-$ in what follows. As we shall see, their two-valuedness will be finally proved at the end of the derivation.

As also seen in Sec.~\ref{complex}, a simple and natural way to account for this doubling is to use the complex numbers and product. This is the origin of the complex value of the wave function in Quantum Mechanics which allows us to recover the global reversibility of physical laws.

Following Eqs.~(\ref{plusvel}) and (\ref{minusvel}) the differential element ${\rm d}X$ can be decomposed as
\beq
{\rm d}X_{\pm} = v_{\pm} {\rm d}t + {\rm d}\xi_{\pm}
\label{rmdminus}
\eeq
\beq
{\rm d}\xi_{\pm} = \eta_{\pm}(2{\cal D})^{1-1/D_F} {\rm d}t^{1/D_F} = \eta_{\pm} \sqrt{2{\cal D}}{\rm d}t^{1/2}
\label{rmdplus}
\eeq
where the $\eta_{\pm}$'s are dimensionless normalized stochastic variables such that $\langle \eta \rangle =0$, $\langle \eta^2 \rangle =1$ and so on for $Y$ and $Z$.

Then two classical derivatives ${\rm d}_{\pm}/{\rm d}t$ can be defined such that ${\rm d}_{\pm}x(t)/{\rm d}t = v_{\pm}$ and combined to construct the complex derivative operator
\beq
\frac{\widehat{{\rm d}}}{{\rm d} t} = \frac{1}{2} \left(\frac{{\rm d}_+}{{\rm d}t} + \frac{{\rm d}_-}{{\rm d}t}\right) -\frac{{\rm i}}{2}\left(\frac{{\rm d}_+}{{\rm d}t} - \frac{{\rm d}_-}{{\rm d}t}\right).
\label{complderop}
\eeq

This operator is applied to the classical part of the position vector and gives a complex velocity which reads
\beq
{\cal V} = \frac{\widehat{{\rm d}}}{{\rm d} t} x(t) = \frac{v_+ + v_-}{2} - {\rm i} \frac{v_+ - v_-}{2}.  
\label{complexveloc}
\eeq

Having defined the covariant derivative, its expression has be obtained by explicitly calculating its effect on some physical quantity.

Since the fractal dimension is 2, the derivative of a scalar function $f$ must be calculated up to second order
\beq
\frac{{\rm d}f}{{\rm d}t} = \frac{\partial f}{\partial t} + \frac{\partial f}{\partial X}\frac{{\rm d}X}{{\rm d}t} + \frac{1}{2}\frac{\partial^2 f}{\partial X^2}\frac{{\rm d}X^2}{{\rm d}t}.
\label{deroff}
\eeq
The stochastic mean of this formula gives, in the second rhs term, $\langle {\rm d} \xi \rangle = 0$, hence $\langle {\rm d}X \rangle = {\rm d}x$ and in the third rhs term, $\langle {\rm d} \xi^2 \rangle = \langle {\rm d} X^2 \rangle = 2 {\cal D} \; {\rm d}t$ with its non fractal part, ${\rm d}x^2/{\rm d}t$, being negligible. Therefore, in three dimensions, 
\beq
\frac{{\rm d}_{\pm}f}{{\rm d}t} = \left(\frac{\partial}{\partial t} + v_{\pm}.\nabla \pm {\cal D} \Delta \right) \, f.
\label{covder}
\eeq
The last step is to recombine the two derivatives into a complex covariant time derivative, such as
\beq
\frac{\widehat{{\rm d}}}{{\rm d} t} = \frac{\partial}{\partial t} + {\cal V}. \nabla - {\rm i} {\cal D}. \Delta
\label{covtimeder}
\eeq
Then, standard mechanics is generalized by using the scale relativistic covariance. First, a complex Lagrange function ${\cal L}(x^i, {\cal V}, t)$, with $i= 1,2,3$ denoting the three space coordinates is defined, and, from it, a complex action ${\cal S} = \int_{t_1}^{t_2}{\cal L}(x^i, {\cal V}, t) \; {\rm d} t$.

Then, a stationary action (geodesic) principle is applied and generalized Euler-Lagrange equations where the covariant derivative $\widehat{{\rm d}}/{\rm d}t$ replaces the full derivative ${\rm d}/{\rm d}t$ are obtained.

Since only the classical parts of the variables are considered (their fractal parts have been taken into account by the covariant derivative), the basic symmetries of classical physics hold: from usual space homogeneity a generalized complex momentum is obtained: ${\cal P} = \partial {\cal L}/ \partial {\cal V}$. Considering the action as a function of the upper integration limit, the action variation from a trajectory to another nearby one yields ${\cal P} = \nabla {\cal S}$. Generalizing the Lagrange function of standard Newtonian mechanics for a closed system to ${\cal L}(x^i, {\cal V}, t) = \frac{1}{2}m{\cal V}^2 - \Phi$ generalized Euler-Lagrange equations keeping the Newtonian form are obtained as,
\beq
m \frac{\widehat{\rm d}}{{\rm d}t} {\cal V} = -\nabla \Phi.
\label{genELeqs}
\eeq

Then, a complex wave function is introduced, which is another expression for the complex action and can therefore be written
\beq
\psi = {\rm e} ^{{\rm i} {\cal S}/S_0}.
\label{defwavefunc}
\eeq
Since ${\cal P} = \nabla {\cal S}$ and $S_0 = \hbar$ in the micro-physics case, ${\cal P} \psi = -{\rm i} \hbar \nabla \psi$ and the correspondence $\widehat{P} = -{\rm i} \hbar \nabla$ is obtained. In the Newtonian case, ${\cal P} = m {\cal V}$, and thus ${\cal V} = \nabla {\cal S} / m = -{\rm i} \; (S_0/m) \, \nabla \ln \psi$.

The fundamental equation of dynamics can now be written in terms of $\psi$, i.e.,
\beq
{\rm i} S_0 \frac{\widehat{\rm d}}{{\rm d}t}(\nabla \ln \psi) = \nabla \Phi.
\label{eqdyn}
\eeq
After some calculations \cite{LN93}, the full equation becomes a gradient which can be integrated as
\beq
{\cal D}^2 \Delta \psi + {\rm i} {\cal D} \frac{\partial}{\partial t} \psi - \frac{\Phi}{2m} \psi = 0 
\label{schrofin}
\eeq

The usual Schr\"odinger equation of Quantum Mechanics corresponds to ${\cal D} = \hbar / 2m$. In other cases, ${\cal D} \neq \hbar / 2m$ characterizes the system as a self-organization constant. 

We have now at our disposal all the tools needed to finally prove the two-valuedness of the classical part of the velocity field. Indeed, the wave function $\psi$ can be written as $\psi=\sqrt{P} \times e^{i \theta}$. This yields, in addition to the geodesic / Euler-Lagrange  (Eq.~\ref{genELeqs}) and Schr\"odinger (Eq.~\ref{schrofin}) representations of the motion equations, a third representation as a fluid-like system of equations of the Euler and continuity type \cite{LN09}. This is a mixed representation with regard to the previous ones, written in terms of the real part $V=2 {\cal D} \nabla \theta$ of the complex velocity field $\cal V$ and of the squared modulus $P = | \psi |^2$ of the wave function. The real and imaginary parts of the Schr\"odinger equation respectively yield a Euler-like equation (after differentiation) and a continuity equation,
\begin{equation}
 \l(\frac{\partial}{\partial t} + V \cdot \nabla\r) V  = -\nabla \l(\frac{\phi+Q}{m}\r), \;\;\;\;\; \frac{\partial P}{\partial t} + {\rm div}(P V) = 0,
\end{equation}
where an additional  potential energy $Q$ has emerged, that reads
\begin{equation}
\label{Q1}
Q =-2m{\cal D}^2 \frac{\Delta \sqrt{P}}{\sqrt{P}}. 
\end{equation}
This final form of the equations of the velocity field of geodesics finally ensures the Born interpretation according to which $P= |\psi|^2$ \cite{NC07}.

Then the two classical velocity fields can now be calculed from these variables. One finds
\beq
v_+= V+{\cal D} \nabla \ln P, \;\;\;\;\; v_-= V -{\cal D} \nabla \ln P,
\eeq
and they are therefore equal only in the case $P=$cst, i.e. only in the fully classical case. In all other cases, these relations definitely prove that there is generally a two-valuedness, not only of the fractal, divergent part of the velocity field, but also of its classical (differentiable) part. 

Before concluding, let us illustrate this result by an explicit example. In the case of the $n=2$, $l=1$, $m=1$ orbital of the hydrogen atom, one finds in spherical coordinates , $V_r=0$, $V_\theta=0$, $V_\varphi=  \csc\theta/r$, $U_r= (1/r)-(1/2)$, $U_\theta=\cot \theta /r$, $U_\varphi=0$, and therefore the two velocities $v_+=V+U$ and $v_-=V-U$ are indeed different, since one obtains
\beq
v_+= \l(  \frac{1}{r} - \frac{1}{2}, \; \frac{ \cot \theta}{r}, \; \frac{ \csc \theta}{r}  \r), \;\;\; v_-= \l( - \frac{1}{r} + \frac{1}{2}, \;- \frac{ \cot \theta}{r}, \; \frac{ \csc \theta}{r}  \r).
\eeq


\section{Conclusions}
\label{concl}

In this paper, we have attempted to make clearer the nature of the scale variables in the theory of Scale Relativity, and to analyze in more detail their effect on the fundamental laws of motion. A continuous but non-differentiable space (more generally space-time) is fractal, which means that its geometric description involves an explicit dependence on scale variables.

From the viewpoint of the theoretical description, the differential elements which one makes tending to zero are the fundamental scale variables. The possibility to define these differentials is a direct consequence of the {\em continuity} of the space-time manifold. Therefore the non-differentiability manifests itself, not by making us unable to differentiate, but by the fact that ratios of differential elements, like ${\rm d} x/{\rm d} t$, no longer exist at the limit ${\rm d} t \to 0$

The Scale Relativity method solves this problem by simply taking into account all what happens when ${\rm d} t \to 0$, but without taking necessarily the limit. In other words, $v={\rm d}X /{\rm d} t$ is considered as an explicit function $v({\rm d} t)$ of ${\rm d} t$. When the limit ${\rm d} t \to 0$ does exist, this is the differentiable case, and it is included as a particular case of the new differential calculus. When it does not exist, the new method uses still a description tool in terms of the function $v({\rm d} t)$, and it simply means that $v({\rm d} t) \to \infty$ when ${\rm d} t \to 0$.

Another fundamental scale variable results from the fact that our access to physical phenomena is always made through measurements, and that a measurement apparatus is always characterized by a measurement resolution $\varepsilon >0$.  We have shown in the present work that the fundamental two-valuedness of velocities (time-derivatives) which gives rise to the complex nature of the wave function in the scale-relativistic foundation of quantum mechanics is fundamentally issued from the relation between these two scale variables, ${\rm d} t$ and $\varepsilon_t$. 

As regards obtaining a Schr\"odinger equation as a prime integral of the geodesic equation (more generally, of the fundamental equation of dynamics) under fractality conditions, one could ask why it does not apply to usual diffusion, for example in the case of the physical Brownian motion of submillimeter particles. The answer is that, strictly, the theory assumes fractality of space-time without any lower limit to the scale dependence that fractality and continuity implies. Such a space-time divergence is expected to be achieved only at scales smaller than the deBroglie length of a given system, i.e., it corresponds only to usual quantum mechanics. However, we have suggested that the theory could also be applicable, now in an approximate way, to fractal macroscopic systems. But it is clear that this application would be valid only provided two conditions:  (i) The range of scales between the upper effective de Broglie scale and the lower cut-off of the fractal behavior  should be large enough for an effective constant fractal dimension to be established; we have estimated that this ratio should be $\gg 10^3$ for the theory to be applicable \cite[Chap.~10]{LN11}. (ii) The dynamics should be Newtonian (i.e., a force creates an acceleration), while mesoscopic diffusive systems, such as the physical motion of Brownian sub-micrometer particles, are often characterized by a Langevin-like dynamic (the force is proportional to velocity); in this case, one does not obtain a Schr\"odinger-type equation \cite{LN11}. These conditions may nevertheless be fulfilled in astrophysical large scale systems, such as protoplanetary nebulae, in which planetesimals show a chaotic motion due to gravitational scattering by the other bodies of the planetary disk, and in which the range of scales is very large and the dynamics remains Newtonian \cite{LN93,LN97,NSL00,DRN03}. 

We will, in a companion paper, generalize these results to space differentials and to the (motion-)relativistic case which lead to bi-spinorial (bi-quaternionic) wave function solutions of the Klein-Gordon and Dirac equations \cite{CN13}.

Moreover, it has been shown that the fundamental nature of scale variables is tensorial \cite{LN11} and this fact plays a central role in the application of the scale relativity approach to gauge fields \cite{NCL06}.  A generalization of the present results to this case will be made in a forthcoming work. \\


{\it Acknowledgments}. One of the authors (LN) acknowledges helpful remarks and questions by Dr. Hans Dembinski, the answer to which has led to the present paper.



\end{document}